\documentclass[aps,prl,twocolumn,superscriptaddress]{revtex4}
\usepackage{graphicx}
\usepackage{color}
\usepackage{amsmath}
\usepackage{braket}
\usepackage{mhchem}

\begin{document}

\title{Charge density wave-generated Fermi surfaces in NdTe$_3$}

\author{Alla Chikina}
\affiliation{Department of Physics and Astronomy, Interdisciplinary Nanoscience Center, Aarhus University, 8000 Aarhus C, Denmark}

\author{Henriette Lund}
\affiliation{Department of Physics and Astronomy, Interdisciplinary Nanoscience Center, Aarhus University, 8000 Aarhus C, Denmark}

\author{Marco Bianchi}
\affiliation{Department of Physics and Astronomy, Interdisciplinary Nanoscience Center, Aarhus University, 8000 Aarhus C, Denmark}

\author{Davide Curcio}
\affiliation{Department of Physics and Astronomy, Interdisciplinary Nanoscience Center, Aarhus University, 8000 Aarhus C, Denmark}

\author{Kirstine J. Dalgaard}
\affiliation{Department of Chemistry, Princeton University, Princeton, New Jersey 08544, USA}

\author{Martin Bremholm}
\affiliation{Department of Chemistry, Interdisciplinary Nanoscience Center, Aarhus University, 8000 Aarhus C, Denmark}

\author{Shiming Lei}
\affiliation{Department of Chemistry, Princeton University, Princeton, New Jersey 08544, USA}

\author{Ratnadwip Singha}
\affiliation{Department of Chemistry, Princeton University, Princeton, New Jersey 08544, USA}

\author{Leslie M. Schoop}
\affiliation{Department of Chemistry, Princeton University, Princeton, New Jersey 08544, USA}

\author{Philip Hofmann}
\email{philip@phys.au.dk}
\affiliation{Department of Physics and Astronomy, Interdisciplinary Nanoscience Center, Aarhus University, 8000 Aarhus C, Denmark}
\date{\today}
\begin{abstract}
The electronic structure of NdTe$_3$ in the charge density wave phase (CDW) is investigated by angle-resolved photoemission spectroscopy.  The combination of high-quality crystals and careful surface preparation reveals subtle and previously unobserved details in the Fermi surface topology, allowing an interpretation of the rich and unexplained quantum oscillations in the rare earth tritellurides RTe$_3$. In particular, several closed Fermi surface elements can be observed that are related to CDW-induced replicas of the original bands, leading to the curious situation in which a CDW does not only remove Fermi surface elements but creates new ones that are observable in transport experiments. Moreover, a large residual  Fermi surface is found in the CDW gap, very close to the position of the gapped normal-state Fermi surface. Its area agrees very well with high-frequency quantum oscillations in NdTe$_3$ and its presence is explained by strong electron-phonon coupling combined with the quasi one-dimensional character of the CDW. Finally, we identify the origin of the  low-frequency $\alpha$ quantum oscillations ubiquitous for the lighter R elements in the RTe$_3$ family and responsible for the high mobility in these compounds.
\end{abstract}
\maketitle

The rare-earth tritelluride compounds \ce{RTe3} are prototypical charge density wave (CDW) compounds with a quasi one-dimensional (1D) CDW emerging on a square lattice of Te atoms. They have been studied as model systems for CDW physics but also because of their magnetic properties, the competition of CDW transitions and magnetism and the analogy of the CDW to ordering phenomena in high-$T_c$ cuprates -- but without the difficulty of strong correlations  \cite{Yumigeta:2021aa}. Recent years have seen a renewed interest in \ce{RTe3} because of fascinating possibilities to modify the electronic ground state by magnetic fields, ultrashort light pulses, or mechanical stress \cite{Schmitt:2008aa,Rettig:2016aa,Zong:2018ab,Kogar:2019ab,Walmsley:2020aa,Dalgaard:2020aa,Straquadine:2022tv}, and as a model system to study the Higgs (amplitude) mode of the CDW by quantum interference methods \cite{Wang:2022tq}. 

On the other hand, the rich spectrum of quantum oscillations (QOs) in the \ce{RTe3} materials  \cite{Ru:2008ac,Sinchenko:2016tr,Walmsley:2020aa,Dalgaard:2020aa,Watanabe:2021vf} is poorly understood. Some frequencies have been assigned to Fermi surface (FS) features that are not expected to be affected by the CDW. Several QOs, however, have not been assigned to any FS element, notably the low-frequency $\alpha$ QOs that are ubiquitous for light R \ce{RTe3} and thought to be responsible for the high mobility \cite{Lei:2020aa,Dalgaard:2020aa}. Moreover, recent studies of de Haas–van Alphen and  Shubnikov-de Haas QOs in high mobility crystals have revealed high-frequency QOs that should originate from large FS elements, covering a substantial fraction of the Brillouin zone (BZ) \cite{Dalgaard:2020aa,Lei:2020aa}. However, there are no obvious candidates for such large FS elements in the CDW phase, since the CDW is expected to lead to a large gap in most of the BZ, as has been confirmed by angle-resolved photoemission spectroscopy (ARPES) in numerous studies \cite{Gweon:1998tt,Komoda:2004aa,Brouet:2004ab,Schmitt:2008aa,Brouet:2008aa,Moore:2010ab,Schmitt:2011aa,Rettig:2014aa,Rettig:2016aa,Lee:2016vn,Lei:2020aa}. The assignment of the most of the observed QOs to FS elements is still an outstanding problem.

We thus revisit the electronic structure of \ce{NdTe3} by ARPES, using the same high mobility samples that show the high-frequency QOs  \cite{Dalgaard:2020aa}. Combined with a careful surface preparation that avoids ever exposing the samples to air, we obtain ARPES results revealing not only detailed fine structure of band interactions and bilayer splitting but also FS elements inside the CDW gap and FS pockets that are created by CDW-induced replica bands. These FS elements can explain most of the observed QOs but challenge the conventional picture of CDWs as primarily removing FS elements. 

\begin{figure}
  \includegraphics[width=0.5\textwidth]{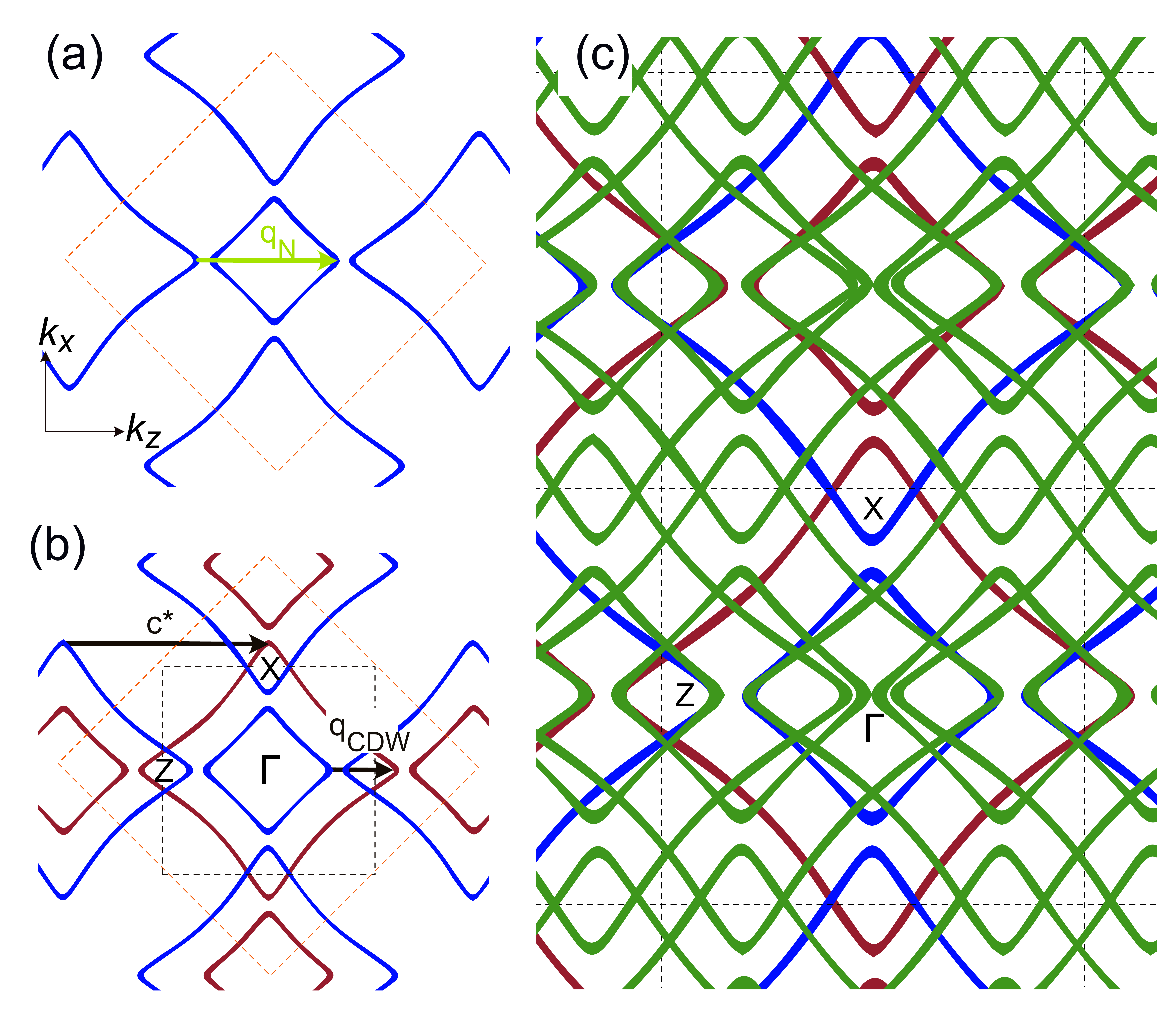}
  \caption{(Color online) Simplified sketch of the \ce{RTe3} Fermi surface (FS)  and the CDW formation. (a) FS of a single square net of Te atoms. The dashed line is the Brillouin zone and $\mathbf{q}_\mathrm{N}$ is the nesting vector. (b) The three-dimensional structure of \ce{RTe3} results in a smaller BZ (black dashed line and high-symmetry points) and a back-folded FS (red). In this situation, the CDW nesting vector is $\mathbf{q}_\mathrm{CDW}$. (c) Extended zone view of the situation in panel (b) with all the shadow bands (green), created by displacing all the FS features by $\pm \mathbf{q}_{CDW}$.
  }
  \label{fig:1}
\end{figure}

The FS of \ce{RTe3} arises from interactions between quasi 1D $p_x$ and $p_z$ bands in a square lattice of Te atoms. An excellent discussion of this can be found in, e.g., Refs. \cite{Brouet:2004ab,Brouet:2008aa}. The Appendix of this paper also contains a detailed introduction.
In Fig. \ref{fig:1}, we recapitulate the main features. Fig. \ref{fig:1}(a) shows the Fermi contour for a single layer of Te atoms when no CDW is present. It arises from the  $p_x$ and $p_z$ bands with avoided crossings, creating the ``square'' FS that lies completely inside the first BZ, and four ``outer'' pockets crossing the BZ boundary (here, the nomenclature of FS elements and tight-binding parameters follows  Refs. \cite{Voit:2000aa,Brouet:2004ab,Brouet:2008aa}). The instability towards CDW formation arises from possible interactions between the outer FSs and the square via the nesting vector  $\mathbf{q}_{\mathrm{N}}$. In addition, due to the three-dimensional crystal structure of  \ce{RTe3}, the unit cell projected onto the Te sheets is larger than the unit cell of the square net, leading to a smaller BZ and thus to a back-folding of bands and FS elements into this smaller BZ. Both BZs are shown in Fig. \ref{fig:1}(b), along with the additional FS elements created by back-folding of the original FS into the smaller BZ (in red). These FS elements are consequently referred to as the ``folded'' FS. As far as the CDW is concerned, the introduction of a smaller BZ requires a description by a shorter nesting vector $\mathbf{q}_\mathrm{CDW} = \mathbf{c}^{\ast} - \mathbf{q}_\mathrm{N}$ between the original and the folded FS, where $ \mathbf{c}^{\ast} $ is the reciprocal lattice vector of the large unit cell in the $z$ direction. Introducing a periodicity corresponding to $\mathbf{q}_\mathrm{CDW}$  creates replicas of the original bands shifted by $\pm \mathbf{q}_\mathrm{CDW}$ (plus higher orders). These are the green so-called CDW shadow bands in  Fig. \ref{fig:1}(c). Interaction between the shadow bands and the original bands then opens the CDW gap around  $k_x = 0$.  
We stress that this description of the CDW mechanism is strongly simplified. Indeed, it is clear that other factors such as strong and momentum-dependent electron-phonon coupling contribute  \cite{Johannes:2008aa,Eiter:2012aa,Zhu:2015ac,Maschek:2015uk}.
As a final detail, the structure of \ce{RTe3} contains two layers of Te atoms adjacent to each other and an interaction between these layers gives rise to an additional small splitting of the FSs known as bilayer splitting \cite{Gweon:1998tt,Brouet:2004ab,Laverock:2005aa,Brouet:2008aa}.

\ce{NdTe3} crystals were grown using the same procedure as in Ref. \cite{Dalgaard:2020aa}. To avoid crystal degradation due to air exposure, a glove box was used for opening the crystal ampules and for mounting the samples on the holders used for ARPES. From the glove box, the samples  were moved into an ultra-high vacuum (UHV) suitcase and transferred to the ARPES setup.  The sample were then cleaved in UHV prior to measurement.
ARPES data for several samples were collected at the SGM-3 beamline of ASTRID2 \cite{Hoffmann:2004aa} at a temperature of 35~K, with an energy resolution varying between 60 and 25~meV and an angular resolution of 0.02$^{\circ}$. All samples showed very nearly the same electronic structure but subtle details changed between samples. We therefore show data from two representative cases.

\begin{figure}
\includegraphics[width=0.5\textwidth]{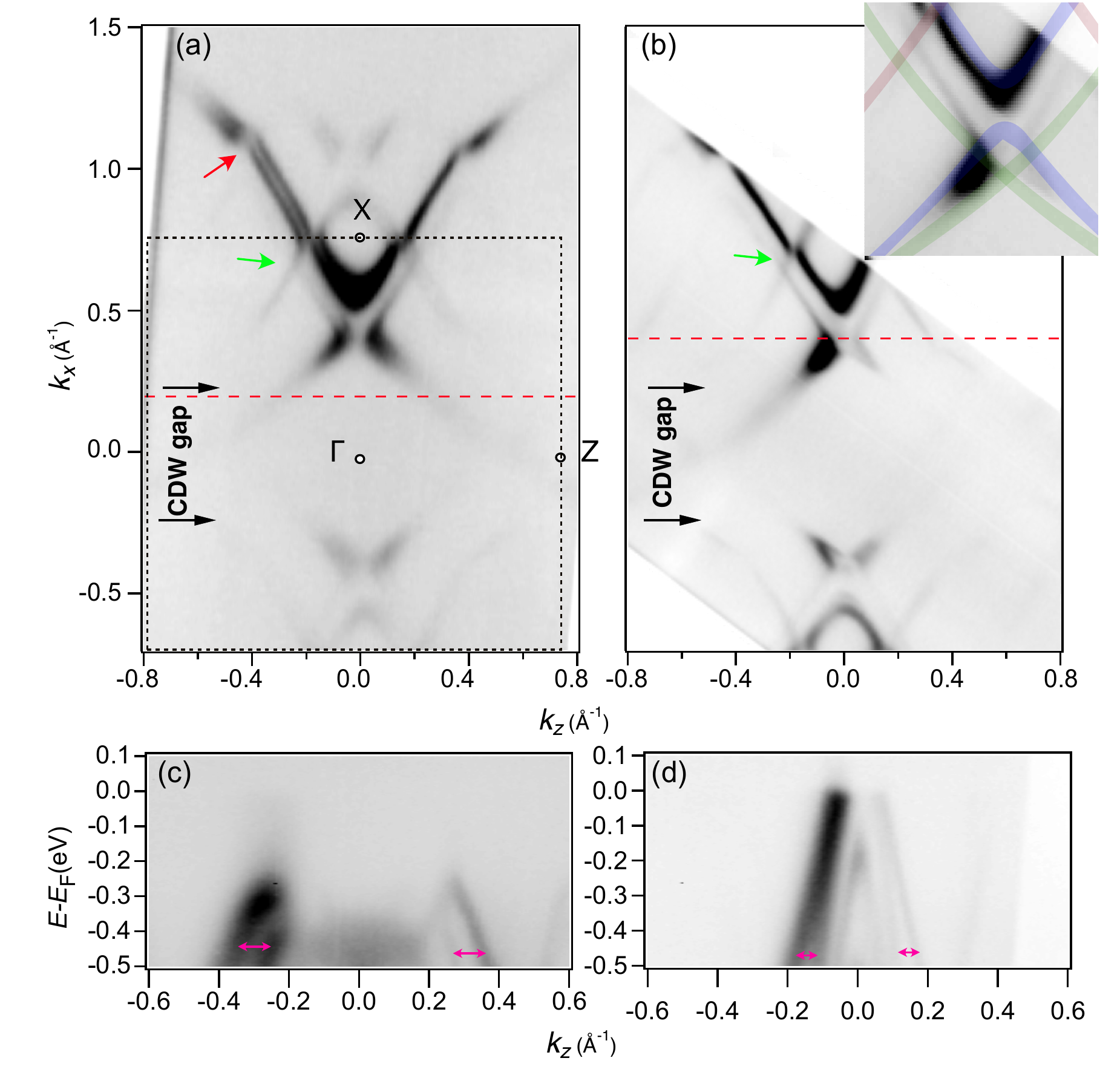}
  \caption{(Color online)  Photoemission intensity at the Fermi level of \ce{NdTe3} (integrated in a $\pm25$~meV window). Dark corresponds to high intensity. (a) and (b) show the results from two different samples, collected at photon energies of 55~eV and 35~eV, respectively. The Brillouin zone and the region gapped by the CDW are indicated. The green arrows show the crossing of the folded band and the CDW shadow band. The red arrow shows an avoided crossing between the outer FS and the CDW shadow band. The inset in (b) is a magnification of the situation around the X point with the simple sketch of the bands from Fig. \ref{fig:1}(c) superimposed. (c) and (d) Photoemission intensity along the dashed lines in (a) and (b), respectively.  The arrows indicate the bilayer splitting between some of the bands. 
  }
  \label{fig:2}
\end{figure}

Starting with an overview of the Fermi surface topology, the photoemission intensity at the Fermi level of \ce{NdTe3} is given in Fig. \ref{fig:2}(a) and (b) for two different samples. The maps are generally in agreement with previously reported ARPES studies of \ce{RTe3} crystals \cite{Gweon:1998tt,Brouet:2004ab,Komoda:2004aa,Brouet:2008aa,Schmitt:2008aa,Moore:2010ab,Schmitt:2011aa,Rettig:2014aa,Rettig:2016aa,Lee:2016vn}, showing closed FS elements around the X-point, created by joining the original and the folded FS, and a gap opening (removal of the FS) in a large fraction of the BZ around  $k_x=0$.

Compared to previous ARPES studies of \ce{RTe3}, the features in the maps are sharper and we can observe more details.
In Fig. \ref{fig:2}(a), a clear doubling of the FS elements appears around the X point and of the outer FS branch. This is ascribed to bilayer splitting. 
For the second sample in Fig. \ref{fig:2}(b), the bilayer splitting is not observed at the FS but other features appear sharper. The bilayer splitting can be detected for higher binding energies in both samples  \cite{Brouet:2008aa}, as shown by the cuts in Fig. \ref{fig:2}(c) and (d) that are taken along the red dashed lines in panels (a) and (b), respectively. Pairs of bilayer-split bands are indicated by arrows in panels (c) and (d). Fig. \ref{fig:2}(c) shows data inside the CDW gap, illustrating the opening of a large gap. 
There is no detectable avoided crossing between the CDW shadow band and the folded band in Fig. \ref{fig:2}(a) and (b) (marked by green arrows) in either sample, confirming earlier results \cite{Brouet:2008aa} and underlining the layer-confined character  of the CDW. Finally,  new features observed here are the disruption of the outer FS in panel (a) (red arrow), indicating an avoided crossing between the main band and the CDW shadow band (see Fig. \ref{fig:1}(c)) and the fine structure around the un-gapped corner of the inner square FS. These are in reasonably good agreement with a tight-binding model, as shown in the Appendix.

We now attempt an understanding of the full FS topology, trying to reconcile the ARPES results with those from recent QO experiments. In short, Ref. \cite{Dalgaard:2020aa} reports five main frequencies, $\alpha$, $\beta_{1,2}$, $\gamma_{1,2}$, $\delta_{1,2}$ and a weak QO $\eta$. The presence of $\alpha$, $\beta_{1,2}$ and $\gamma_{1,2}$ frequencies agrees with an investigation of \ce{NdTe3} in Ref. \cite{Walmsley:2020aa}  but the high-frequency $\eta$ and $\delta_{1,2}$ structures had not been reported before (note that the nomenclature is different between Refs. \cite{Walmsley:2020aa} and \cite{Dalgaard:2020aa} and we adopt the latter).

When expressing the observed QO frequencies in units of the BZ area, defined by $a^{\ast}c^{\ast}$ =2.09~\AA$^{-2}$, where $a=c= 4.35$~\AA, their areas correspond to:   $\beta_{1,2}$ = 2.1\% and 2.3\%;  $\gamma_{1,2}$ = 3.7\% and 3.9\%.  The weak high-frequency QO  $\eta$ corresponds to the 9.6\% and the largest frequencies $\delta_{1,2}$ to 16.6 and 17.5\%. The structure called $\alpha$  actually contains several frequencies, the dominant of 0.2\%, as well as two additional ones of 0.7\% and 0.9\% (see Ref.  \cite{Dalgaard:2020aa} incl. SM and in particular Fig. S10).

\begin{figure}
\includegraphics[width=0.5\textwidth]{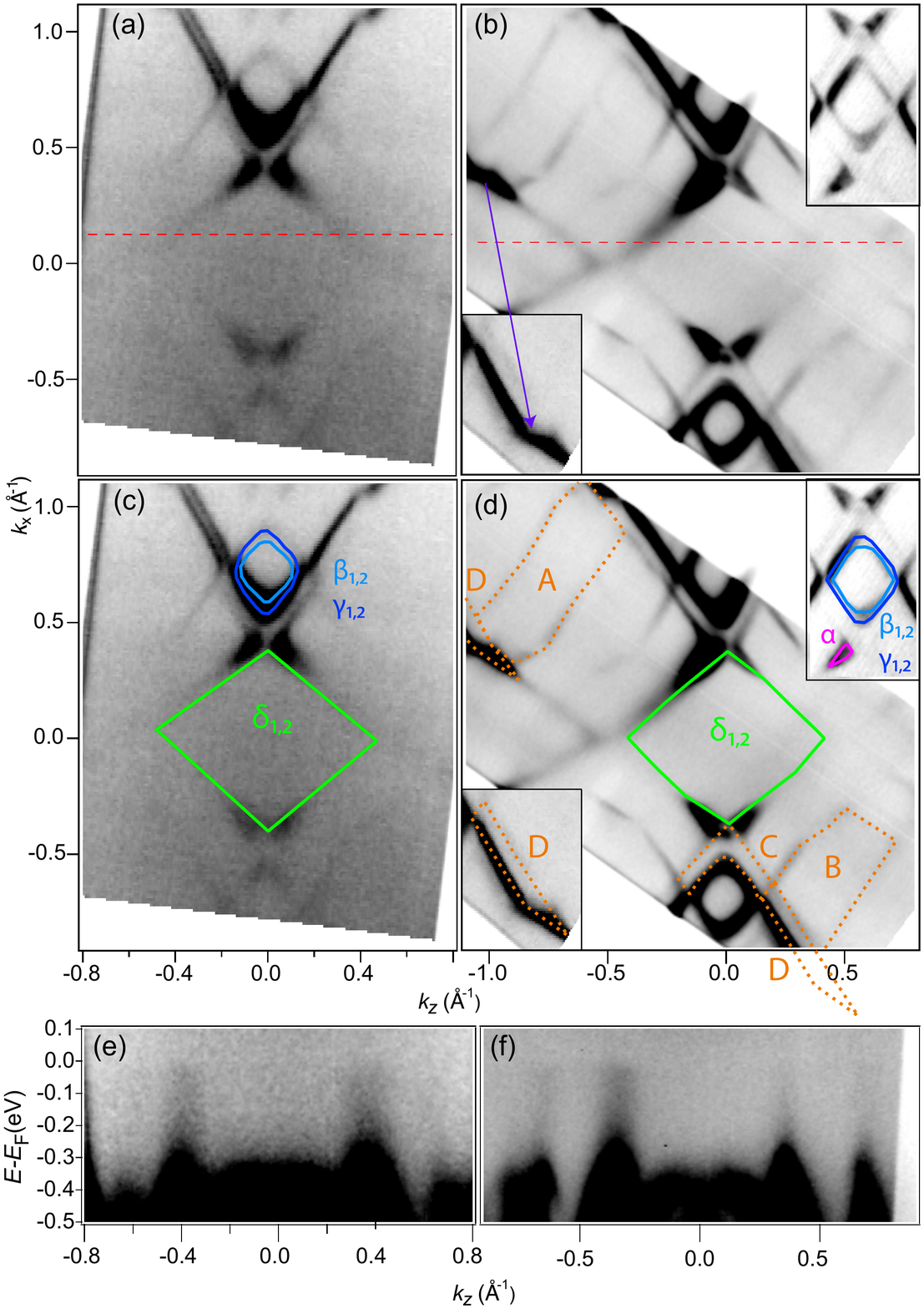}\\
  \caption{(Color online)  (a), (b) Same data as in Fig. \ref{fig:2}(a) and (b) but with the greyscale saturated to emphasise the weaker features. The upper inset shows the ``butterfly'' structure near X collected in a higher BZ (integrated in an energy window of $\pm$5~meV around $E_F$). The lower inset displays entire structure marked by an arrow (only partly visible in the panel). (c), (d) Same as (a) and (b) but with superimposed closed FS elements for smallest $\alpha$, $\beta_{1,2}$, $\gamma_{1,2}$ and $\delta_{1,2}$  structures, as well as for elements involving shadow bands (A-D, dashed lines). (e), (f) Photoemission intensity along the dashed lines in (a) and (b), respectively. Again, the greyscale is saturated for most of the features. 
  }
  \label{fig:3}
\end{figure}

The approach to assigning the observed frequencies to the ARPES data is illustrated in Fig. \ref{fig:3}. Fig.  \ref{fig:3}(a) and (b) show the same data as Fig. \ref{fig:2}(a) and (b) but the greyscale is chosen such that many features are saturated, to make the weaker structures visible. The upper inset of Fig. \ref{fig:3}(b) shows a high-resolution view of the FS near  the X point taken from a higher BZ, giving better $k$-resolution. Panels (c) and (d) show the same data as (a) and (b) but with a tentative assignment of the Fermi contours superimposed as colored outlines. These were obtained by a fit to momentum distribution curves near the Fermi level. For the contours within the CDW gap, this was possible due to the presence of weak un-gapped bands crossing $E_\mathrm{F}$, as will be discussed below.    

The intermediate  $\beta_{1,2}$ and $\gamma_{1,2}$ QOs have already been assigned to the pockets around the X point in Ref. \cite{Ru:2008ac} and this is supported by our data. The areas of the contours averaged over Fig. \ref{fig:3}(c) and (d) are 2.2$\pm$0.1 and 3.1$\pm$0.1\% of the BZ, respectively, in good agreement with the QO results. Note that the splitting of the $\beta$ and $\gamma$  QOs can be explained by  a small $k_y$ dispersion for each bilayer-split FS tube.

The high-resolution data near the X point reveals a tiny pocket in the ``wings'' of the ``butterfly'' structure and we assign this to the lowest frequency $\alpha$ pocket observed for the entire \ce{RTe3} family with light R elements \cite{Ru:2008ac,Walmsley:2020aa,Dalgaard:2020aa,Watanabe:2021vf}. This pocket has not been identified previously in ARPES.  We estimate the area to be 0.16$\pm$0.1\% of the BZ. The fact that the wings form a closed contour is seen in the inset of Fig. \ref{fig:3}(b) and in the individual energy vs $k$-cuts through the structure in Fig. \ref{fig:s2}. It is challenging to determine how the individual bands contribute to the butterfly structure. As seen in the inset of Fig. \ref{fig:2}(b), we expect contributions from the main $p_x$ and $p_z$ bands and the CDW-shifted shadow bands in this region and these four bands are again split by the bilayer interaction.

The CDW-gap in a large part of the BZ makes it difficult to identify a FS element that could be responsible for the $\delta_{1,2}$ QOs, simply because of their size. We assign these frequencies to a closed contour similar to the inner part of the FS surface, as shown in Fig. \ref{fig:3}(c) and (d). 
Although this structure resides almost entirely within the CDW gap, there are some compelling arguments for this assignment.
First of all, a weak shadow of this FS is always present in the maps of Fig. \ref{fig:2} and \ref{fig:3}. This shadow is created by a residual intensity near the location of the inner FS. Its origin is seen in Fig. \ref{fig:3}(e) and (f) which shows the photoemission intensity along lines deep inside the gap with the greyscale saturated such that weak features are visible. Very weak un-gapped bands dispersing up to the Fermi level are observed \emph{in addition to the gapped states}. These bands are responsible for the observed photoemission intensity at the Fermi level in the gap and they appear to form a weak FS corresponding to what would be expected for the material in the normal state. The pocket formed by this  FS has an area of 17.3$\pm$0.2\% of the BZ, averaged over the two data sets and in excellent agreement with the $\delta_{1,2}$ frequencies reported from QOs.  We note that one also needs to consider the possibility that the in-gap FS elements could be created by some type of experimental artefacts, such as crystal twinning or detector non-linearities \cite{Reber:2014td}. As discussed in the Appendix, these scenarios can be ruled out.

The most likely explanation for the presence of this pocket is that the electron-phonon coupling opens the CDW gap but leaves the spectral function  inside the gap finite. In 1D, this is expected for a CDW driven by electron-phonon coupling \cite{Lee:1973up,Zhao:2005wh}, and while the Te-net in \ce{RTe3} is two-dimensional (2D), the CDW shows several characteristics of being 1D \cite{Sinchenko:2012wj}.  It is interesting that such a situation can lead to QOs despite  the fact that there is no real Fermi level crossing, just a finite spectral function at $E_F$, a situation that has been invoked previously to describe QOs in a pair density wave state of the cuprate superconductors \cite{Norman:2018uq}. Strong and $q$-dependent electron-phonon coupling has repeatedly been discussed as the driving mechanism of CDWs in \ce{RTe3} \cite{Johannes:2008aa,Zhu:2015ac} and experimentally shown to be present in the lattice vibrations \cite{Eiter:2012aa,Maschek:2015uk}. Our results are consistent with this and establish the manifestation and significant consequences of electron-phonon coupling in the electronic structure.

An alternative explanation for the finite intensity in the gap is  an electronic phase separation between CDW and normal state regions of the sample, leaving some minority part of the sample in the normal state. This could be induced either by defects \cite{Zhang:2014wg} or by imperfect nesting \cite{Rice:1970vh}. Electronic phase separation has been observed in other CDW systems, e.g., in 1T-\ce{TiSe2}  \cite{Jaouen:2019wf},  \ce{Lu2Ir3Si5} \cite{Lee:2011wl} or indium atomic wires on Si \cite{Zhang:2014wg}. In the case of \ce{RTe3}, however, published scanning tunnelling microscopy results show no indications of phase separation \cite{Ralevic:2016aa,Fu:2016ws,Fang:2019vp,Fang:2007wh}. 

We note that the in-gap photoemission intensity observed here is consistent with previous ARPES studies of \ce{RTe3} materials \cite{Schmitt:2008aa,Moore:2010ab,Lee:2016vn,Zong:2018ab}. However, the phenomenon was not discussed.

Finally, the weak $\eta$ as well as the two remaining $\alpha$ frequencies need to be accounted for.  ARPES reveals several possible candidates for closed contours matching these. These are outlined in Fig. \ref{fig:3}(d) as dashed lines and labelled A-D, with sizes of A  (10.8$\pm$0.3\%),  B (8.0$\pm$0.3\%), C (2.4$\pm$0.2\%), D (1.4$\pm$0.2\%), such that the A and D contours can tentatively be assigned to the $\eta$ and highest $\alpha$ QO. Interestingly, these contours are not present in the original FS without the CDW and arise only due to the presence of CDW-induced shadow band crossings. In fact, there potentially is a multitude of new closed FS contours induced in this way (see Fig. \ref{fig:1}(c)) but not all of these are observed, as the shadow bands are generally weak in ARPES \cite{Voit:2000aa,Brouet:2004ab,Brouet:2008aa}.

To the best of our knowledge, QOs arising from CDW shadow bands and residual normal state FSs have not been reported before. Their presence raises interesting questions about the energy balance of the CDW. A nesting-driven CDW is commonly viewed as electronically stabilised by a partial removal of the FS \cite{Gruner:1994aa}. While this picture is too simple to explain many real systems including \ce{RTe3} \cite{Laverock:2005aa,Johannes:2008aa,Eiter:2012aa,Zhu:2015ac,Maschek:2015uk}, it is intriguing that the CDW formation should not only remove FS elements but also create new ones, as this leads to an electronic energy \emph{increase}. A proper account for the CDW's energy balance thus requires considering the many-body spectral function of the CDW state, especially for the incommensurate case, for which the conventional band structure picture is, strictly speaking, not valid \cite{Zhang:2015ae}.

In conclusion, we have addressed the open question of assigning QOs in the \ce{RTe3} group to specific FS contours. This has revealed the origin of the $\alpha$ pocket that is responsible for the high mobility in the materials. We have also found several QOs that can be explained by CDW-induced shadow bands. This leads to the interesting scenario in which the introduction of a CDW not only removes FS elements but also creates new ones, relevant for the transport properties and the energetics of the CDW state. Finally, we have found a remaining spectral weight with the shape of the original FS in the gap of the CDW which we have assigned to the high-frequency QO observed for \ce{NdTe3} and \ce{GdTe3}. This can be seen as a manifestation of the 1D character of the CDW, where a finite spectral function in the gap is expected for an electron-phonon coupling driven CDW. 

\begin{acknowledgments}
This work was supported by VILLUM FONDEN via the Centre of Excellence for Dirac Materials (Grant No. 11744). Work at Princeton was supported by the Gordon and Betty Moore Foundation's EPIQS initiative (grant number GBMF9064) and the Arnold and Mabel Beckman foundation through a BYI grant awarded to LMS. We thank Veronique Brouet, Jennifer Cano and Rafael Fernandes for stimulating discussions.
\end{acknowledgments}

\section{Appendix}

In this appendix material, we give a more detailed introduction into the FS topology for \ce{RTe3}. The discussion repeats parts of the what is written in the main paper for better readability. We provide additional experimental data on the $\alpha$ pocket; we discuss  experimental situations that could potentially give rise to  artefact-induced weak bands inside the CDW gap, and we present an extended version of the tight-binding model introduced in Refs. \cite{Brouet:2004ab,Brouet:2008aa}.

\subsection{The Fermi Surface of \ce{RTe3}}

\begin{figure}
  \includegraphics[width=0.5\textwidth]{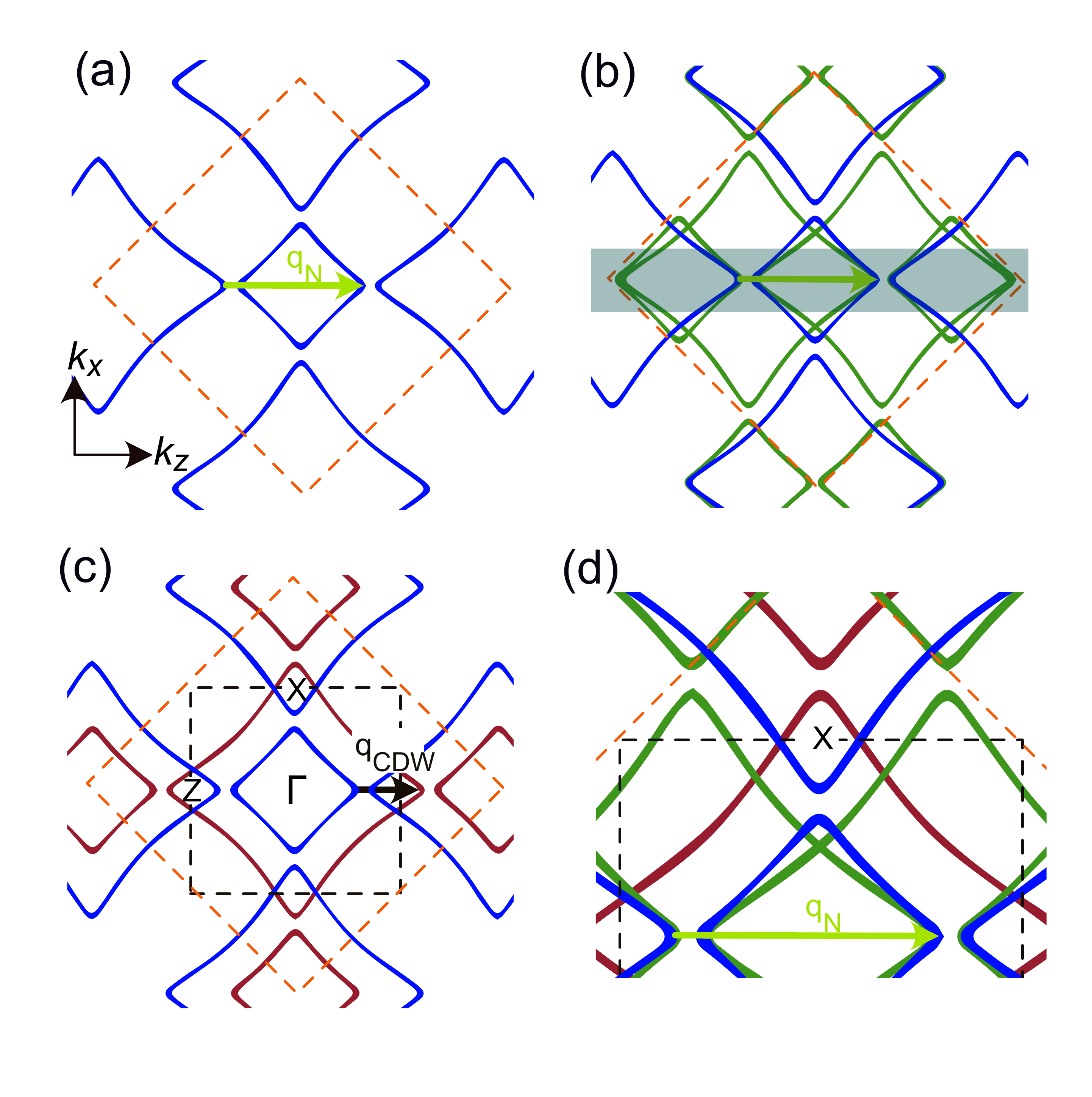}
  \caption{Simplified sketch of the \ce{RTe3} FS  and the CDW formation. (a) FS of a single square sheet of Te atoms, resulting from nearly one-dimensional $p_x$ and $p_y$ bands, combined with an avoided crossing at their intersection. The dashed line is the Brillouin zone (BZ) boundary and $\mathbf{q}_\mathrm{N}$ is the nesting vector. (b) Original FS (blue) and CDW-induced shadow FS elements (green), obtained by shifting the original FS by $\pm \mathbf{q}_\mathrm{N}$. The interaction between the original and the shifted FS elements leads to a gap opening in the shaded region. (c) The three-dimensional structure of \ce{RTe3} results in a smaller BZ (black dashed line and high-symmetry points) and a back-folded FS (red). In this situation, the CDW gap can be created by displacing the FS elements by $\pm \mathbf{q}_{CDW}$. (d) Close-up of the situation near the X point point with the original, shadow and folded FS elements indicated. 
  }
  \label{fig:s1}
\end{figure}

Compared to the main text, we give a somewhat more extended discussion of the main FS features that are important for the quantum oscillations in \ce{RTe3} with a single CDW \cite{Ru:2008ac,Sinchenko:2016tr,Walmsley:2020aa,Dalgaard:2020aa,Watanabe:2021vf}. 
The FS of \ce{RTe3} arises from interactions between quasi one-dimensional orthogonal $p_x$ and $p_z$ states in a square lattice of Te atoms, combined with the fact that there are four such lattices in the three-dimensional unit cell. Fig. \ref{fig:1}(a) shows the quasi two-dimensional Fermi contour for a single layer of Te atoms when no CDW is present.  Fig. \ref{fig:s1}(a) reproduces the sub-figure here for easier readability. The nearly one-dimensional diagonal lines from the $p_x$ and $p_z$ dispersion are discernible with interruptions by avoided crossings, creating the ``square'' FS that lies completely inside the first Brillouin zone (BZ) and four ``outer'' pockets crossing the BZ boundary (our nomenclature of FS elements and tight-binding parameters follows mostly Refs. \cite{Voit:2000aa,Brouet:2004ab,Brouet:2008aa}). The instability towards CDW formation arises from interactions between the outer FS and the square. To see this, suppose that a new periodicity $2\pi / q_{\mathrm{N}}$ is introduced into the system along the $z$ direction. This would be expected to create replicas of the electronic structure displaced by the vector $\pm \mathbf{q}_\mathrm{N}$, as shown in Fig. \ref{fig:s1}(b) for displacements of the original FS by  $\pm \mathbf{q}_\mathrm{N}$. The shifted FS overlaps with the original FS in a wide range around $k_x = 0$, a property called nesting. An interaction between the the overlapping FS elements then results in a gap opening around the Fermi level and an electronic energy gain, stabilising the CDW state. The green, shifted FS elements are often referred to as ``shadow'' FS. We stress that this description of the CDW mechanism is strongly simplified. Indeed, it is not even such that (only) nesting is driving the CDW  \cite{Laverock:2005aa,Johannes:2008aa,Eiter:2012aa,Zhu:2015ac,Maschek:2015uk}.

Due to the three-dimensional (3D) crystal structure of  \ce{NdTe3}, the actual FS topology is more complicated than that in Fig. \ref{fig:s1}(a). The effective unit cell parallel to the Te sheets is larger, leading to a smaller BZ and thus to a back-folding of bands and FS elements into this smaller BZ. Both BZs are shown in Fig. \ref{fig:s1}(c), along with the additional red FS elements created by the back-folding of the original FS into the smaller BZ. These red FS elements are consequently referred to as the ``folded'' FS and they play an important role in understanding the quantum oscillations in the materials. The folded bands can be obtained by shifting the original blue bands by a reciprocal lattice vector of the larger 3D lattice, $\mathbf{c}^{\ast}$, in the $z$ direction.  As far as the CDW is concerned, the introduction of a smaller BZ requires a description by a shorter nesting vector $\mathbf{q}_\mathrm{CDW} = \mathbf{c}^{\ast} - \mathbf{q}_\mathrm{N}$ between the original and the folded FS. This description is equivalent to the one using $\mathbf{q}_{\mathrm{N}}$ in Fig. \ref{fig:s1}(a) and an overall gap around $k_x = 0$ can again be obtained by shifting both the original and the folded FS by $\pm\mathbf{q}_\mathrm{CDW}$.  However, when viewing the CDW as an essentially two-dimensional phenomenon, it is an appropriate simplification to introduce the CDW only in the original bands by a shift of  $\pm \mathbf{q}_\mathrm{N}$ as in Fig. \ref{fig:s1}(b). Fig. \ref{fig:s1}(d) shows this simplified model that we will use in the tight-binding calculations below. It contains the main bands, the folded bands and the main bands displaced by $\pm \mathbf{q}_\mathrm{N}$. Note that this model does not have the correct periodicity expected for smaller 3D BZ (even though it is sufficient for introducing a gap in the 1st BZ). The correct periodicity is only obtained when using $\mathbf{q}_\mathrm{CDW}$ and the result is shown in Fig. \ref{fig:1}(c) of the main text. 

\subsection{Supplementary experimental data on the $\alpha$ pocket}

\begin{figure}
\includegraphics[width=0.45\textwidth]{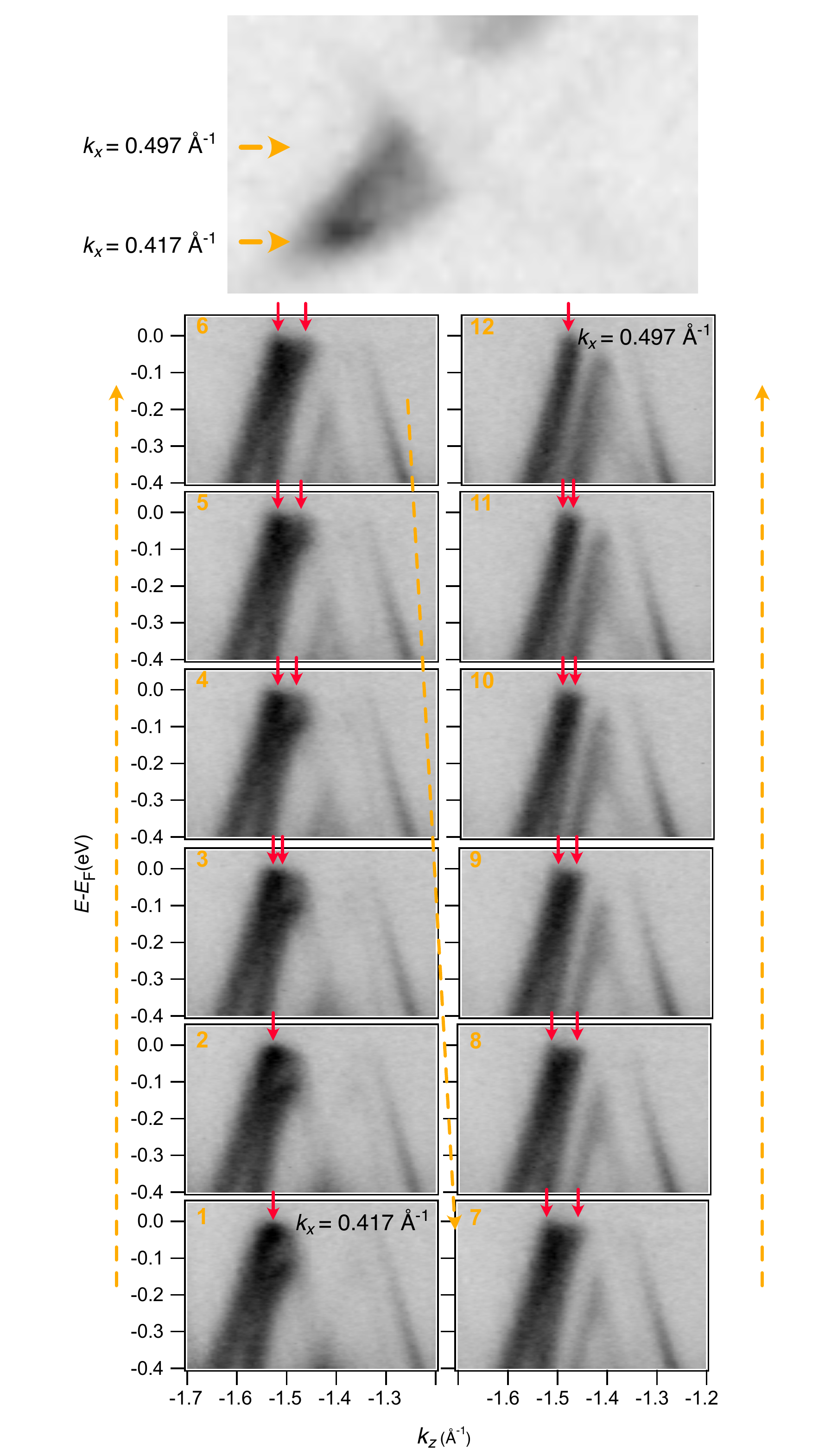}\\
  \caption{Row-wise photoemission intensity through the butterfly wing (shown on top) with red arrows indicating the position of Fermi level crossings. The dashed arrows and the numbers give the ordering of the images.
  }
  \label{fig:s2}
\end{figure}

Fig. \ref{fig:s2} illustrates the closed pocket character of the butterfly wing, assigned to the lowest frequency $\alpha$ QO in the main text, by presenting row-wise spectra through the wing (shown on the top) between the two arrows. The arrow(s) in each panel show an estimate of the Fermi level crossing(s), demonstrating the separation of a single Fermi level crossing into two and the re-joining of the two crossings into one, forming the closed pocket. 

\subsection{Experimental artefacts capable of creating weak bands in ARPES}

\begin{figure}
\includegraphics[width=0.35\textwidth]{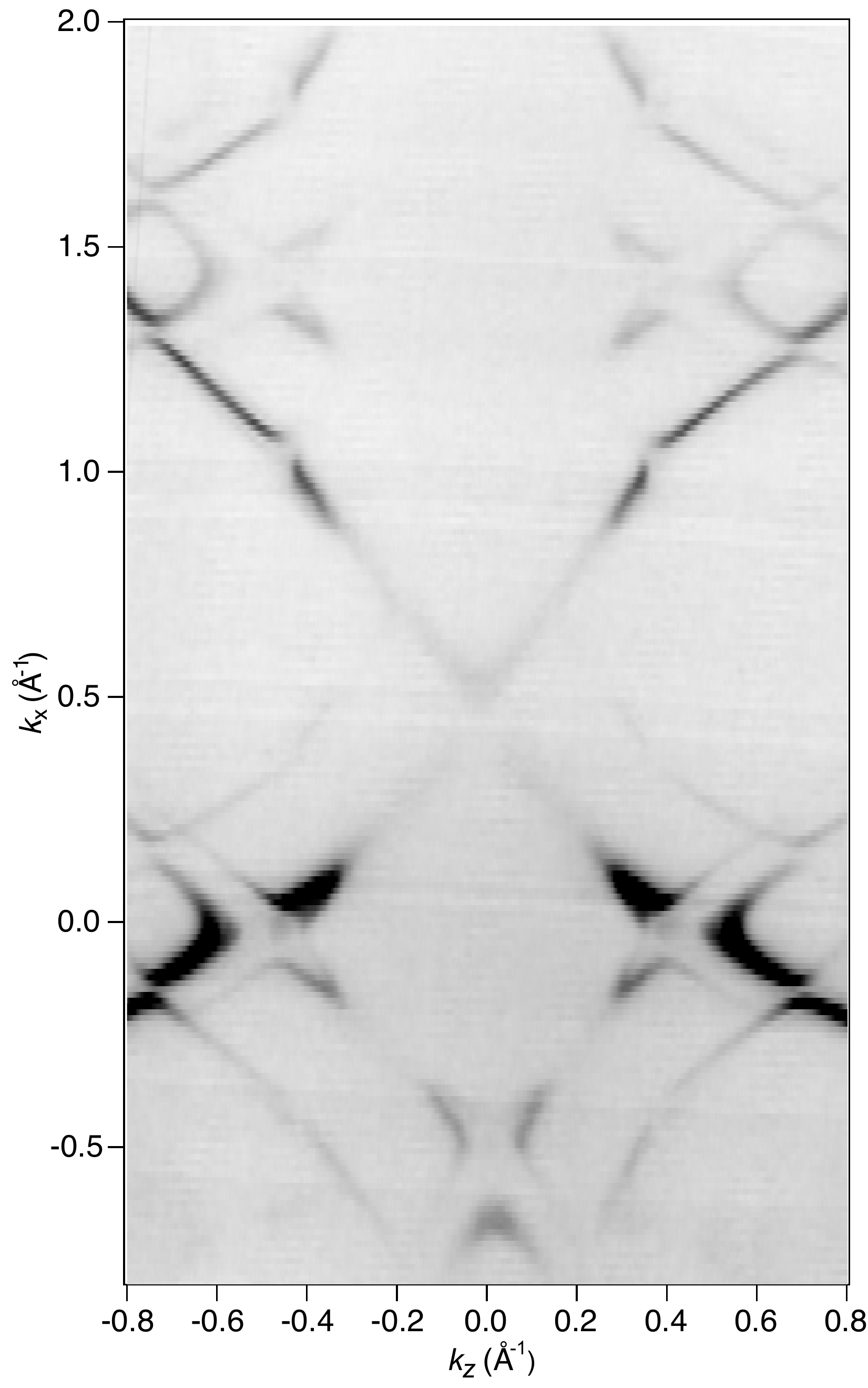}\\
  \caption{Data corresponding to Figs. \ref{fig:2} and \ref{fig:3} (photoemission intensity at the Fermi level) but for a sample location with a substantial contribution from a mirror-twin orientation domain. 
  }
  \label{fig:s3}
\end{figure}

One needs to consider the possibility that the weak bands forming the  FS inside the gap could be created by some type of experimental artefact. The most obvious possibility arises from the possibility that the crystals can be twinned, i.e., they can contain macroscopic domains in which the $a$ and $c$ directions are interchanged. It appears thus conceivable that the observed in-gap FS actually originates from a small contamination of twin domain within the area probed by ARPES, as the twin-domain CDW gap opens in the orthogonal direction. When scanning the light spot across the surface of some samples, we could in fact observe a superposition of twin domains in the spectra, for an example see Fig. \ref{fig:s3}.
 Based on such data, we can exclude that the residual FS is caused by a twin domain using two arguments: (1) The strongest feature of the inner FS is always the butterfly structure near the X point, independent of the twin domain orientation with respect to the light polarisation (see Fig. \ref{fig:s3}).  The observed weak FS in Figs. \ref{fig:2} and \ref{fig:3} can thus not be caused by a twin domain because this would  simultaneously give rise to a pronounced butterfly structure in the $\Gamma$-Z direction of the main domain. Such a structure is not observed in Figs. \ref{fig:2} and \ref{fig:3}. (2) The CDW-induced gap around the $\Gamma$-Z axis is very wide, at least 0.25~\AA$^{-1}$, consistent with previous observations \cite{Brouet:2008aa}. Very large parts of the observed weak FS would lie in that gap, even for a rotated twin domain being simultaneously present, also ruling out a twin domain as the origin of the weak FS. 
 
Another possible cause of weak features in ARPES, similar to the residual FS, is the fact that the detector can tend to ``streak out'' structures of particularly high intensity, giving rise to spurious intensity \cite{Reber:2014td}. This is in fact  observed for the most intense features but, being an artefact of the electron detection, such ``streaks'' do not end at the Fermi energy, as the bands giving rise to the weak FS do.

\subsection{Tight-binding model}

\begin{figure}
\includegraphics[width=0.5\textwidth]{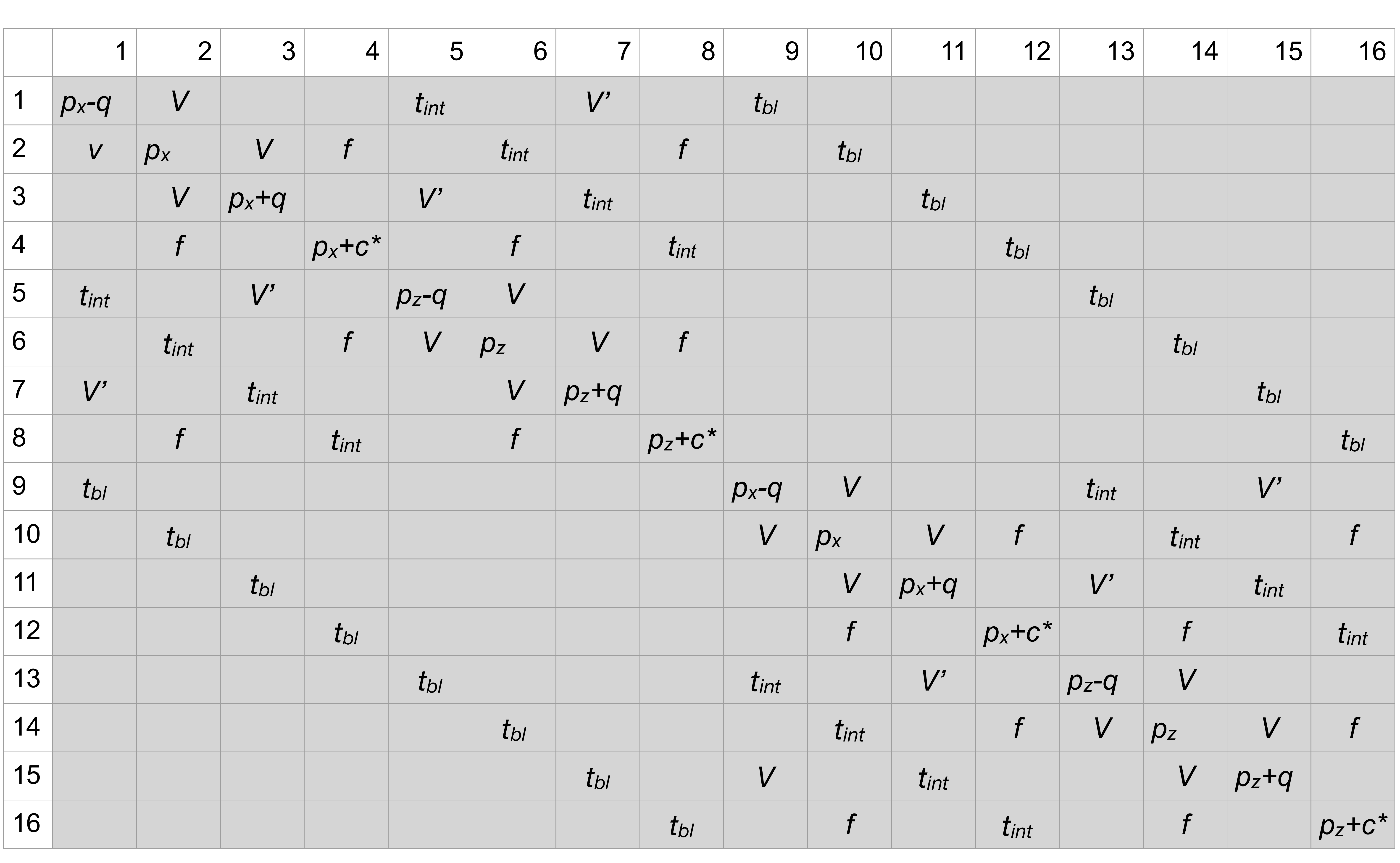}\\
  \caption{Matrix elements for the tight-binding model.
  }
  \label{fig:s4}
\end{figure}

We present an  extended version of the tight-binding model introduced by Brouet \emph{et al.} \cite{Brouet:2008aa} for a qualitative comparison of some experimental features to calculations. This model is based on the original $p_x$ and $p_z$ bands with a dispersion given by
\begin{align}
E_{p_x}=-2t_{\|}\cos[(k_x+k_z)\frac{a}{2}] -2t_{\perp}\cos[(k_x-k_z)\frac{a}{2}]-E_\mathrm{F} \nonumber \\
E_{p_z}=-2t_{\|}\cos[(k_x-k_z)\frac{a}{2}]  -2t_{\perp}\cos[(k_x+k_z)\frac{a}{2}]-E_\mathrm{F}, 
\end{align}
where $t_{\|}$ and $t_{\perp}$ are the parameters controlling the coupling along and perpendicular to the chain of p-orbitals, respectively, and $E_\mathrm{F}$ is the Fermi energy given by $E_\mathrm{F} = -2t_{\|} \sin(\pi / 8)$. In addition to these bands, the model contains the shadow bands, obtained by shifting the original $p_x$ and $p_z$ bands by $\pm \mathbf{q}_N$, and the folded bands, obtained by shifting the original bands by $\mathbf{c}^{\ast}$. This gives rise to an $8 \times 8$ matrix. Bilayer splitting is also included, using an interaction $t_\mathrm{bl}$ between all the bands in the model and their bilayer-split counterparts, increasing the size of the matrix to $16 \times 16$. In total, the model has seven parameters: In addition to $t_{\|}$, $t_{\perp}$ and $t_\mathrm{bl}$, there is the interaction $t_\mathrm{int}$ between the $p_x$ and $p_z$ bands, creating the avoided crossings that separate the FS into the square and the outer part, a parameter $f$ controlling the interaction between the original band and the folded bands and the parameter $V$ controlling the size of the CDW gap. Finally, we include an interaction $V^{\prime}$ between two CDW-shifted shadow bands (green bands in Fig. \ref{fig:s1}(d)). This will turn out to be important to describe the situation close to the crossing of these bands near the corner of the inner FS and the X point. The full matrix  is given in Fig. \ref{fig:s4}. The diagonal elements correspond to the dispersions in equation (1) with the notation that, e.g., $p_x$ corresponds to $E_{p_x}$. 
Table \ref{tab:1} gives the parameters used for the model. For the generation of panels (a) - (f) of Fig. \ref{fig:s5}, some of these parameters have been set to zero (see below).

\begin{table}
\caption{\label{tab:1}Parameters of the tight-binding model.} 
\begin{ruledtabular}
\begin{tabular}{p{2.5cm} p{2.5cm} } 
Parameter & Value \\
$a$ & 4.35~\AA\\
$c$ & 4.35~\AA\\
$q_\mathrm{N}$ & 0.68$c^{\ast}$\\
$t_{\|}$ & -1.5~eV\\
$t_{\perp}$ & 0.3~eV\\
$t_\mathrm{int}$ & 0.2~eV\\
$f$ & 0.15~eV\\
$V$ & 0.35~eV\\
$V^{\prime}$ & 0.16~eV\\
$t_\mathrm{bl}$ & 0.035~eV\\
\end{tabular} 
\end{ruledtabular} 
\end{table}

When plotting the results of the model, we do not show all the 8 (16) solutions but only the squared weight coefficients for the original $p_x$ and $p_z$ orbitals, including bilayer splitting \cite{Voit:2000aa,Brouet:2004ab,Brouet:2008aa,Moore:2010ab}. The result is broadened, simulating a Gaussian energy and $k$-resolution function of 25~meV and 0.02~\AA$^{-1}$ full width at half maximum, respectively.

Note that this is a minimal model with some limitations: The CDW is created by a shift of $\pm \mathbf{q}_\mathrm{N}$ between the original $p_x$ and $p_z$ bands in a single layer of Te atoms, and the three-dimensional character of the problem is ignored for simplicity (using the three dimensional description would require a displacement by $\mathbf{q}_\mathrm{CDW}$, creating twice as many bands). As a consequence, there are no shadow bands that gap the red, folded FS in Fig. \ref{fig:s1}(c) and (d) around $k_x = 0$. In fact, the model does not have the full periodicity of the folded band structure, as can be seen be comparing the FS in the two adjacent ``lower'' and ``upper'' BZs in Fig. \ref{fig:s1}(d) because it does not include the CDW shadow bands of the folded bands. None of this matters for the inspection of the CDW and the original bands in the 1st BZ and the situation around the X point, which is the main purpose of using the tight-binding model here. 

\begin{figure*}
 \includegraphics[width=0.9\textwidth]{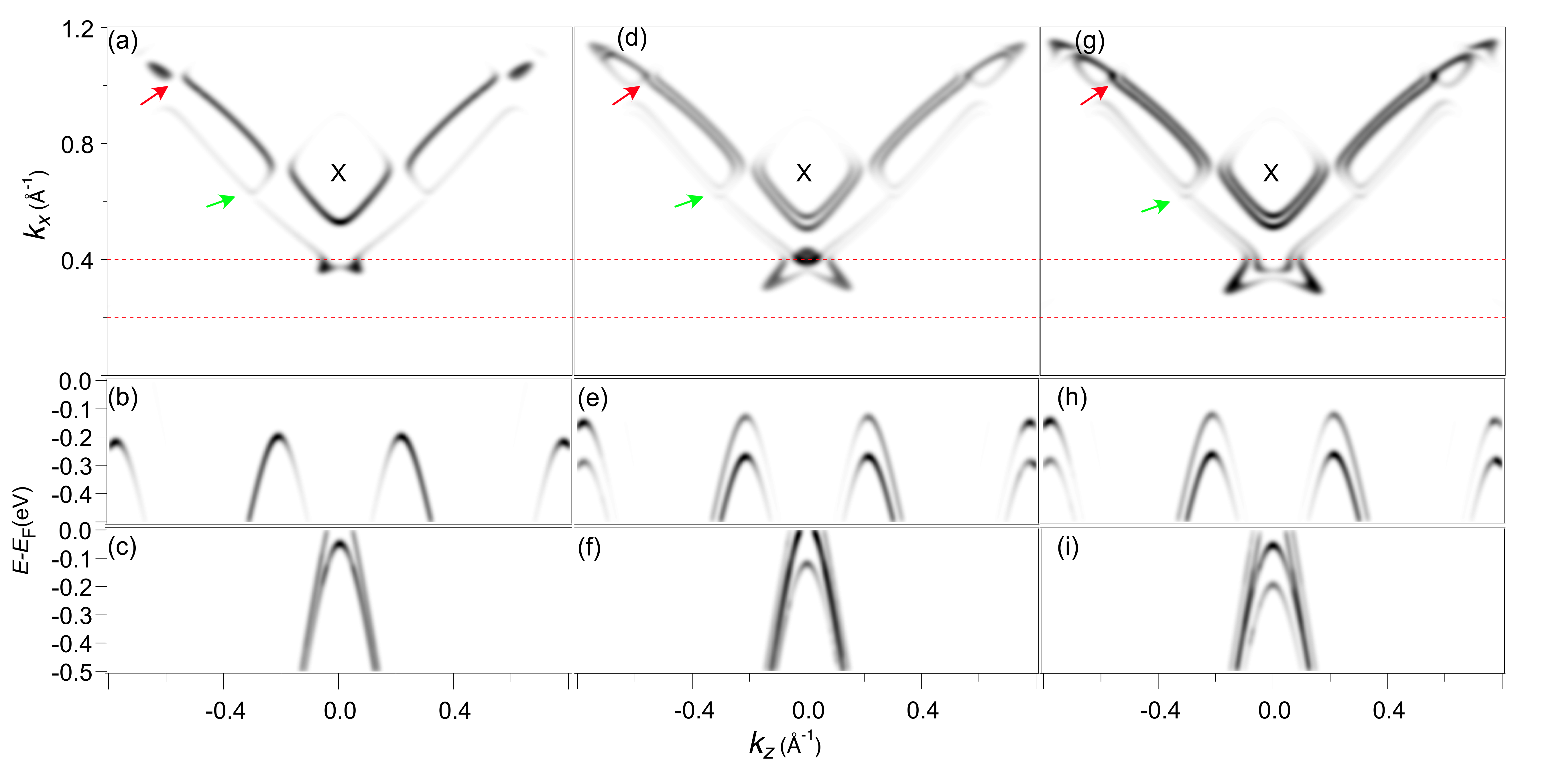}\\
  \caption{Results of the tight-binding model, given as the resolution-broadened spectral weight on the original bands. (a)-(c) Fermi surface without bilayer splitting ($t_\mathrm{bl}=0$) and interaction between the CDW shadow bands ($V^{\prime}=0$). (a) FS,  (b) and (c) Cuts as a function of energy and $k_z$ for $k_x = 0.4$~\AA~and $k_x = 0.2$~\AA, respectively. (d)-(f) Corresponding but with $t_\mathrm{bl}=0.035$~eV and $V^{\prime}=0$. (g)-(i) Corresponding but with $t_\mathrm{bl}=0.035$~eV and $V^{\prime}=0.16$~eV.
  }
  \label{fig:s5}
\end{figure*}

The tight-binding model can explain some features in the data that are new compared to the results already discussed by Brouet \emph{et al.} \cite{Brouet:2004ab,Brouet:2008aa}. Fig. \ref{fig:s5}(a)-(c) show the tight-binding results as a function of $k_x$ and $k_z$  integrated in a $\pm25$~meV window around the Fermi level, as well as two cuts as a function of energy and $k_z$, for $k_x = 0.4$~\AA$^{-1}$ and $k_x = 0.2$~\AA$^{-1}$, respectively. Bilayer splitting and the $V^{\prime}$ interaction between the shadow bands  are not yet included but many details in the experimental Fermi surface are remarkably well reproduced, including the large CDW gap, as can be seen by comparing Fig. \ref{fig:s5}(b) and Fig. \ref{fig:2}(c) in the main text. Also, the closed contour around the X point is formed. The observed disruption of the outer FS (red arrow in Fig. \ref{fig:2}(a)) is reproduced and also marked by a red arrow in Fig. \ref{fig:s5}(a). According to Fig. \ref{fig:s1}(d), this avoided crossing takes place between the outer FS (blue) and the CDW-shifted shadow FS (green). The situation around the butterfly near the X-point appears to be  well reproduced since the model appears to match the finely resolved details in Fig. \ref{fig:3}(b). However, the agreement is actually not as good as it appears because the inset of Fig. \ref{fig:3}(b) and Fig. \ref{fig:s2} show that the wing of the butterfly is actually a closed contour, giving rise to the $\alpha$ quantum oscillation.  Finally, an interesting detail in the model of Fig. \ref{fig:s5}(a) is that the point of intersection between the CDW shadow band and the folded band (green arrows in Fig. \ref{fig:2} and Fig. \ref{fig:s5}) shows a (very weak) avoided crossing. This is  surprising because the tight-binding model does not contain an interaction between the shadow band and the folded band that would cause this. The origin of the weak anti-crossing must thus be the fact that the neighbouring crossing of the main band and the folded band mixes the character of these two and thereby gives rise to a (very weak) interaction with the shadow band. This is not observed in the experiment. 

Fig. \ref{fig:s5}(d)-(f) show the corresponding results when a small bilayer interaction of $t_\mathrm{bl}=$~35~meV is included. This modification creates the desired bilayer splitting of the FS which is in excellent agreement with the results of Fig. \ref{fig:2}(a), especially for the outer FS and the pockets around X. The cut along $k_x = 0.2$~\AA$^{-1}$ in Fig. \ref{fig:s5}(e) shows that the bands of panel (b) are now strongly split by the bilayer interaction. Note that the qualitative agreement with the experimental FS in the butterfly region is not improved by the introduction of bilayer splitting. If anything, it is worse because the wings of the butterfly are still not closed contours and the pronounced FS element in the centre of the butterfly is not found in the experiment (it is observed at slightly higher binding energy).

This can be improved to some degree by introducing an additional interaction in the model between the two CDW-shifted shadow bands (green bands in Fig. \ref{fig:s2}) as $V^{\prime}=$0.16~eV, leading to the situation shown in Fig. \ref{fig:s5}(g)-(i). The FS topology around the butterfly resembles the experiment more closely and while a closed contour around the butterfly wing has not been achieved either, the bilayer-split bands are quite close to forming one. 

Note that there are limits to what one can expect to obtain from the tight-binding model, especially for the very subtle structures around the butterfly which, after all, are generated by the interaction of no less than eight bands. In fact, the limits of the tight-binding model become already clear when comparing the energy vs. $k$ cuts in Fig. \ref{fig:s5} to the experimental ones in Fig. \ref{fig:2}(c) and (d). It is immediately evident that the tight-binding model shows a much higher spectral weight of the shadow bands near their crossing with the original bands. Moreover, the bilayer splitting is clearly energy-dependent in the experiment, as for example seen in Fig. \ref{fig:2}(d) where it is only observed for the highest binding energies but seems to be very small closer to the Fermi energy. Indeed, a scenario where both split bands flatten out and gradually merge into the corresponding shadow bands, as in Fig. \ref{fig:s5}(e) and (h), is never observed.

Finally, we stress that the tight-binding model in this form does not show the  experimentally observed CDW shadow bands even though they are present in the model. This is due to the fact that their intensity is very strongly suppressed in the plots that only show the weight on the original bands.


\end{document}